\documentclass[12pt,preprint]{aastex}

\begin{document}
\def\ale{\mathrel{\hbox{\rlap{\hbox{\lower4pt\hbox{$\sim$}}}\hbox{$<$}}}}
\def\age{\mathrel{\hbox{\rlap{\hbox{\lower4pt\hbox{$\sim$}}}\hbox{$>$}}}}
\def\msun{\mathrel{$M_\odot$}}

\title{GRB~010921: Strong Limits on an Underlying Supernova from HST}

\author{
P.~A.~Price\altaffilmark{1,2},
S.~R.~Kulkarni\altaffilmark{2}, 
B.~P.~Schmidt\altaffilmark{1},
T.~J.~Galama\altaffilmark{2},
J.~S.~Bloom\altaffilmark{2},
E.~Berger\altaffilmark{2},
D.~A.~Frail\altaffilmark{3,2},
S.~G.~Djorgovski\altaffilmark{2},
D.~W.~Fox\altaffilmark{2},
A.~A.~Henden\altaffilmark{4},
S.~Klose\altaffilmark{5},
F.~A.~Harrison\altaffilmark{2},
D.~E.~Reichart\altaffilmark{2},
R.~Sari\altaffilmark{6},
S.~A.~Yost\altaffilmark{2},
T.~S.~Axelrod\altaffilmark{1},
P.~McCarthy\altaffilmark{7},
J.~Holtzman\altaffilmark{8},
J.~P.~Halpern\altaffilmark{9},
R.~A.~Kimble\altaffilmark{10},
J.~C.~Wheeler\altaffilmark{11},
R.~A.~Chevalier\altaffilmark{12},
K.~Hurley\altaffilmark{13},
G.~R.~Ricker\altaffilmark{14},
E.~Costa\altaffilmark{15},
F.~Frontera\altaffilmark{16,17}
and
L.~Piro\altaffilmark{15}.
}

\altaffiltext{1}{Research School of Astronomy \& Astrophysics, Mount Stromlo
Observatory, via Cotter Road, Weston, ACT, 2611, Australia.}
\altaffiltext{2}{Palomar Observatory, 105-24, California Institute of
Technology, Pasadena, CA 91125.}
\altaffiltext{3}{National Radio Astronomy Observatory, P.O. Box O, Socorro,
NM 87801.}
\altaffiltext{4}{Universities Space Research Association/US Naval
Observatory, Flagstaff Station, P. O. Box 1149, Flagstaff, AZ 86002-1149.}
\altaffiltext{5}{Thuringer Landessternwarte Tautenburg, Karl-Schwarzschild-Observaorium, Sternwarte 5, 07778 Tautenburg, Germany.}
\altaffiltext{6}{Theoretical Astrophysics, 130-33, California Institute
of Technology, Pasadena, CA 91125.}
\altaffiltext{7}{Carnegie Observatories, 813 Santa Barbara Street, Pasadena,
CA 91101.}
\altaffiltext{8}{Department of Astronomy, MSC 4500, New Mexico State
University, P.O.~Box 30001, Las Cruces, NM 88003.}
\altaffiltext{9}{Columbia Astrophysics Laboratory, Columbia University,
550 West 120th Street, New York, NY 10027.}
\altaffiltext{10}{Laboratory for Astronomy and Solar Physics, NASA Goddard
Space Flight Center, Code 681, Greenbelt, MD 20771.}
\altaffiltext{11}{Astronomy Department, University of Texas, Austin, TX 78712.}
\altaffiltext{12}{Department of Astronomy, University of Virginia,
P.O. Box 3818, Charlottesville, VA 22903-0818.}
\altaffiltext{13}{University of California Space Sciences Laboratory,
Berkeley, CA, 94720.}
\altaffiltext{14}{Center for Space Research, Massachusetts Institute of
Technology, Cambridge MA 02139}
\altaffiltext{15}{Istituto Astrofisica Spaziale and Fisica Cosmica, C.N.R.,
Via Gobetti, 101, 40129 Bologna, Italy.}
\altaffiltext{16}{Istituto Astrofisica Spaziale, C.N.R., Area di Tor Vergata,
Via Fosso del Cavaliere 100, 00133 Roma, Italy.}
\altaffiltext{17}{Dipartimento di Fisica, Universita di Ferrara,
Via Paradiso 12, 44100, Ferrara, Italy.}

\begin{abstract}
GRB~010921 was the first HETE-2 GRB to be localized via its afterglow
emission.  The low-redshift of the host galaxy, $z=0.451$, prompted us
to undertake intensive multi-color observations with the {\it Hubble
Space Telescope} with the goal of searching for an underlying
supernova component.  We do not detect any coincident supernova to a
limit 1.34 mag fainter than SN~1998bw at 99.7\% confidence, making
this one of the most sensitive searches for an underlying SN.
Analysis of the afterglow data allow us to infer that the GRB was
situated behind a net extinction (Milky Way and the host galaxy) of
$A_V \sim 1.8$ mag in the observer frame.  Thus, had it not been for
such heavy extinction our data would have allowed us to probe for an
underlying SN with brightness approaching those of more typical Type
Ib/c supernovae.
\end{abstract}


\keywords{gamma rays: bursts}

\section{Introduction}
\label{sec:introduction}

Since the discovery of gamma-ray burst (GRB) afterglows there has been
growing evidence linking GRBs to massive stars: the host galaxies of
GRBs are star-forming galaxies and the position of GRBs appear to
trace the blue light of young stars \citep{bkd02}; some of the host
galaxies appear to be dusty with star-formation rates comparable to
ultra-luminous infrared galaxies \citep{bkf01,fbm+01}. On smaller
spatial scales, there is growing evidence tying GRBs to regions of
high ambient density \citep{gw01,hys+01} and the so-called dark GRBs
arise in or behind regions of high extinction \citep{dfk+01a,pfg+02}.

However, the most direct evidence linking GRBs to massive stars comes
from observations of underlying supernovae (SNe) and X-ray lines.  The
presence of X-ray lines would require a significant amount of matter
on stellar scales (e.g. \citealt{pgg+00}), as may be expected in
models involving the death of massive stars.  However, to date, these
detections (e.g. \citealt{pgg+00,rwo+02}) have not been made with high
significance.

If GRBs do arise from the death of massive stars, then it is
reasonable to expect associated SNe.  The GRB-SN link was
observationally motivated by two discoveries: the association of
GRB~980425 with the peculiar Type Ic SN 1998bw \citep{gvv+98,kfw+98}
and an excess of red light superposed on the rapidly decaying
afterglow of GRB~980326 \citep{bkd+99}.  However, these two
discoveries were not conclusive.  The SN association would require
GRB~980425 to be extra-ordinarily under-energetic as compared to all
other cosmologically located GRBs and the case for GRB 980326 is
weakened by the lack of a redshift for the GRB or the host galaxy.

Nonetheless, the two discoveries motivated searches for similar
underlying SN components.  As summarized in
section~\ref{sec:conclusions}, suggestions of similar red ``bumps'' in
the light curves of various other GRB afterglows have been made (to
varying degrees of confidence).

However, there is little dispute that
the well-studied red bump in the afterglow of GRB~011121 is most easily
explained by an underlying supernova \citep{bkp+02,gsw+02}.
Furthermore, from radio and IR observations of the afterglow
\citep{pbr+02}, there is excellent evidence that the circumburst
medium was inhomogeneous with ambient density $\rho \propto r^{-2}$, as 
expected from a massive star progenitor \citep{cl00}; here, $r$ is the
distance from the progenitor.

These developments are in accordance with the expectation of the
``collapsar'' model \citep{woo93,mw+99}.  In this model, the core of a
rotating massive star collapses to a black hole which then accretes
matter and drives a relativistic jet.  Internal shocks within this jet
first cause bursts of $\gamma$-rays and then subsequently result in
afterglow emission as the jet shocks the ambient medium.

It is important to appreciate that the SN light is primarily powered
by radioactive decay of the freshly synthesized $^{56}$Ni whereas the
burst of $\gamma$-rays are powered by the activity of the central
engine. In the current generation of collapsar models, there is
sufficient flexibility to allow for a large dispersion of $^{56}$Ni
and the energy of the engine. Thus, the next phase of understanding
the GRB-SN connection\footnotemark\footnotetext{A class of models,
known as ``supranova'' models, posit a supernova greatly in advance,
many months, of the the GRB event \citep{vs99}. The long delay was
physically motivated to explain the X-ray lines as arising from a
large spatial region.  The current data (e.g. GRB~011121) do not allow
for such long delays.}  will benefit from (and require) observational
measures of these parameters.

Motivated thus, we have an ongoing program of searches for SNe in GRB
afterglows with the {\it Hubble Space Telescope} (HST).  Here, we
present a systematic search for a SN underlying GRB~010921.  In
\S\ref{sec:observations} we present our observations and the details
of photometry in \S\ref{sec:subphot}.  We fit afterglow models and
constrain the brightness of an underlying SN in
\S\ref{sec:discussion}. We then present an overview of previous such
efforts and conclude in \S\ref{sec:conclusions}.

\section{Observations and Reductions}
\label{sec:observations}

GRB~010921 was detected by the High Energy Transient Explorer (HETE-2)
satellite at 2001 September 21.219 UT \citep{rhl+02} and the position
was refined by the InterPlanetary Network error-box
\citep{hurley+01}. Using the 5-m Hale telescope and the Very Large
Array we discovered the afterglow of this event as well as the
redshift of the host galaxy \citep{pkb+02}.

The low redshift of this event, $z = 0.451$, made it a prime
candidate for a search for an underlying SN.  Accordingly, as a part
of our large {\it Hubble Space Telescope} (HST) Cycle 9 program
(GO-8867, P.~I.: Kulkarni), we triggered a series of observations with
the Wide Field Planetary Camera 2 (WFPC2) aboard HST. Owing to the
lateness in identifying the afterglow candidate, the first observation
was on day 35, slightly after the expected peak of the SN.  At each of
epochs 1--3 we obtained $4 \times 400$~s exposures in each of five
filters (F450W, F555W, F702W, F814W and F850LP) with a single diagonal
dither by 2.5 pixels to recover the under-sampled point-spread function
(PSF).  The fourth epoch was optimized for photometry of the host
galaxy and, accordingly, we increased the exposure time to $4 \times
1100$~s.

We used ``On-The-Fly'' pre-processing to produce debiased, flattened
images.  The images were then drizzled \citep{fh02} onto an image with
pixels smaller than the original by a factor of 0.7 using a {\tt
pixfrac} of 0.8.  After rotation to a common orientation the images
were registered to the first epoch images using the centroids of
common objects in the field.  The typical r.m.s.\ registration errors
were less than 0.15 drizzled pixels.

\section{Host Subtraction and Photometry}
\label{sec:subphot}

The host galaxy of GRB~010921 has an integrated magnitude of $R \sim
22$ mag or about 5$\,\mu$Jy \citep{pkb+02}.  Consequently great care
has to be taken to properly photometer the fading afterglow.  Below,
we review various photometric techniques.

\noindent
{\bf Total magnitudes:} The simplest technique is to perform aperture
photometry (e.g. \citealt{gtv+00,phg+01}).  The afterglow flux
is obtained by subtracting the host flux estimated from a very late time
measurement.  A major concern is that the host flux is dependent upon
the choice of aperture (both center and size).  Thus, if different
images have different seeing then it is possible to obtain an
artificial bump in the light curve.


\noindent
{\bf Host subtraction:} The above concern can be alleviated by
subtracting a late-time image from the earlier images.  The afterglow
may then be easily photometered in the host-subtracted images.  This
method has been used with considerable success by those observing SNe~Ia
(e.g.  \citealt{ssp+98}).

\noindent
{\bf $N\times(N-1)/2$ subtraction:}  In this technique, each image is
subtracted from every other image and the afterglow residual
photometered.  The flux at each epoch can be fit through least-squares,
assuming the flux at the final epoch is zero (Novicki and Tonry,
personal communication).  This method makes use of the fact that the
host galaxy has not been observed only once at late times, but at each
epoch and thus better S/N can be obtained from the over-constrained
system.

We employed the $N\times(N-1)/2$ subtraction technique to photometer
the GRB~010921 afterglow in our HST images.  The images were
subtracted using a modified version of ISIS \citep{alard00} and
photometered using the analytic PSF-fitting routine within Vista
(J.~Tonry, personal communication).  We used the {\tt synphot} package
within IRAF to calculate the response of the instrument and filter
combination to a source with constant flux of 1~mJy; the resulting
values are AB magnitudes \citep{fig+96}, expressed as fluxes.
Corrections were made for charge-transfer (in)efficiency (CTE) using
the prescription of \citet{dolphin00} and aperture-corrected to
infinity.

We have also re-analyzed and photometered ground-based images
\citep{pkb+02,pwh+02} of the afterglow, applying $N\times(N-1)/2$
subtraction.  Since this technique assumes that the flux of the
afterglow in the final epoch is zero, which may not be correct for
these images, we subtracted the appropriate fourth-epoch HST
observation (which we have assumed contains no afterglow) from the
final ground-based images, measured the flux of the afterglow and
added this value to the fluxes derived from the $N\times(N-1)/2$
subtraction.

\begin{figure}[tbp]
\plotone{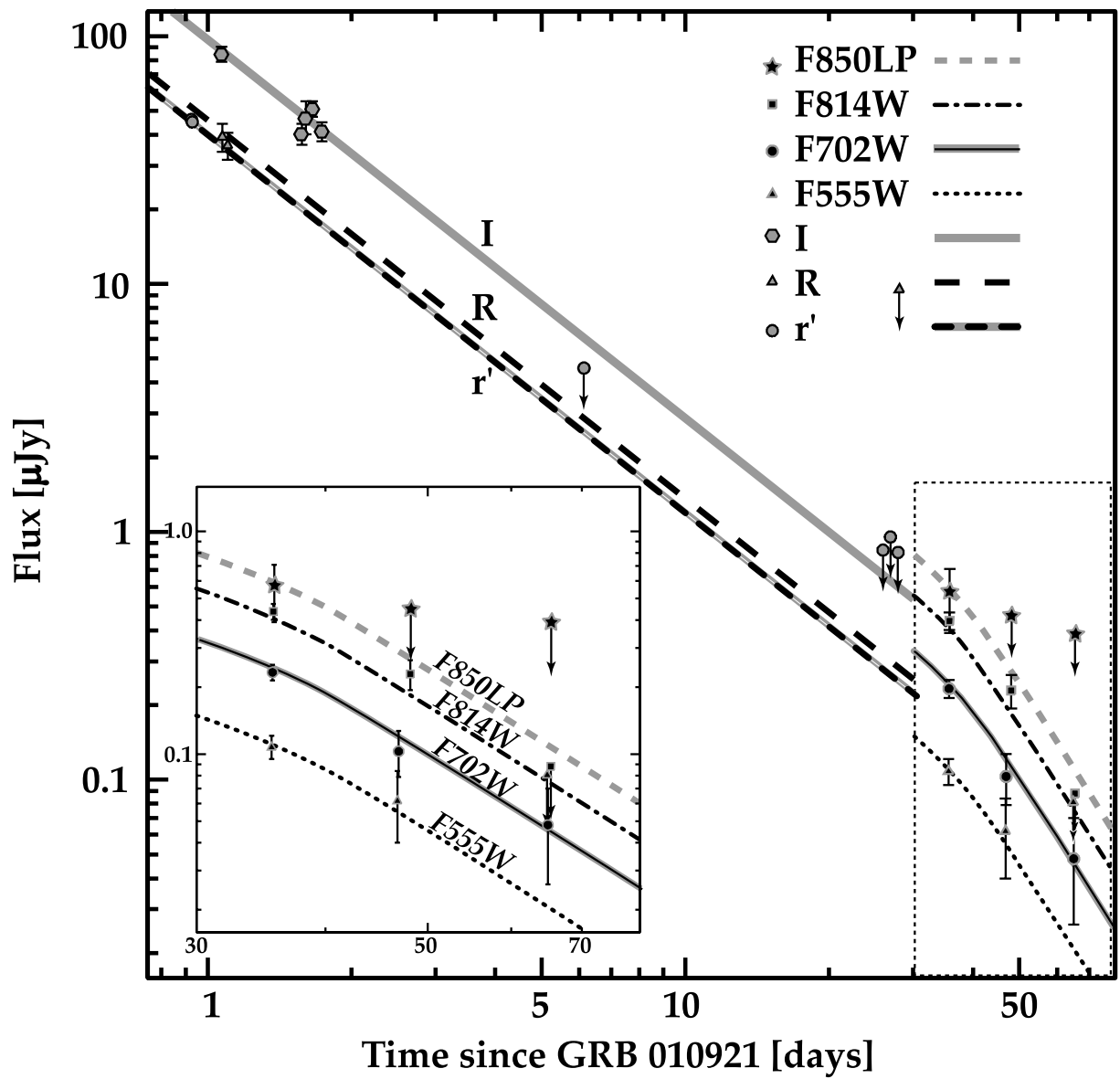}
\caption{The optical light curve of the afterglow of GRB~010921.  Each
data point contains pure afterglow (no contribution from the host
galaxy), and have not been corrected for foreground extinction.
Downward arrows indicate 2$\sigma$ upper limits.  The fit is a
standard broken power-law afterglow model (\S\ref{sec:discussion}).  }
\label{fig:lc}
\end{figure}

The results of the photometry are host-subtracted fluxes for the
afterglow in each of the images, under the assumption that the
afterglow flux in the final HST image (2001 Dec 21) is zero (or 
negligible).  These values are presented in Tables~\ref{tab:hst} and
\ref{tab:ground}.  The values in Table~\ref{tab:ground} supersede the
corresponding measurements presented in \citet{pkb+02} and
\citet{pwh+02}.  We plot the afterglow light curves in
Figure~\ref{fig:lc}.  The light-curves are monotonically decreasing
(i.e. do not level off), and hence we deduce that our assumption of
negligible flux in the final HST image is justified.

\section{Discussion}
\label{sec:discussion}

Temporal breaks in optical light-curves have been seen in many
afterglows and are usually attributed to a ``jet'' geometry (see
\citealt{fks+01}).  We adopt a standard optical afterglow model,
consisting of a broken power-law temporal decay with power-law indices
$\alpha_1$ and $\alpha_2$, and a power-law spectral index, $\beta$
\citep{spn98,sph99}.

Each of these indices are functions of the electron energy
distribution index, $p$, dependent upon the location of the cooling
break relative to the optical bands, and so we consider two cases: the
cooling break is redward of the optical (hereafter, case R); and the cooling
break is blueward of the optical (case B).  We consider, in addition
to a constant circumburst medium, an inhomogeneous circumburst medium,
$\rho\propto r^{-2}$ (see \citealt{cl00} and \citealt{pbr+02}).

We apply the parametric extinction curves of \citet{ccm89} and
\citet{fm88} using the interpolation calculated by \citet{reichart01}.
These extinction curves are characterized by two values, the magnitude
of the extinction in the rest-frame of the host galaxy, $A_V^{\rm
host}$, and the slope of the UV extinction curve, $c_2$ (see
\citealt{reichart01}).  Following \citet{pkb+02}, we adopt $c_2 = 4/3$,
corresponding to an LMC-like extinction curve.  Adopting other
extinction curves (e.g. MW, SMC) yields similar, but more-constraining
results (i.e. any underlying SN must be even fainter than the upper
limit we derive below); see \citet{pkb+02}.

The main purpose of this analysis is to determine whether the
light curves contain an SN component. To this end, we use the
observations of SN~1998bw for an SN template since it is one of the
well observed bright Ib/c SNe which may be related to a GRB
\citep{gvv+98}.  Specifically, we used the $UBVRI$ photometry of
\citet{gvv+98} and derived the flux distribution of SN~1998bw, using
the zero-points and filter curves of \citet{bessell90}.  The resulting
low resolution spectrum (consisting of 5 points at the effective
wavelength of each broadband filter), is redshifted to $z = 0.451$
\citep{pkb+02}, assuming a flat lambda cosmology with $\Omega_{\rm
M}=0.3$ and $H_0 = 65$ km s$^{-1}$ Mpc$^{-1}$.  The redshifted
spectrum, which represents what SN~1998bw would look like at
cosmological distances, is integrated with the appropriate filter
curve to derive the apparent brightness at this redshift.

SN~1998bw at $z = 0.451$ would peak in the rest-frame $I$-band at
approximately 4 $\mu$Jy.  It is evident from Figure~\ref{fig:lc} that
the afterglow is much fainter than this, and, further, that there is
no clear bump in the afterglow light curve.  We therefore allow the SN
component to be scaled by $\delta$ magnitudes in our model.  The SN is
placed behind the same foreground (i.e.\ Milky Way) and host galaxy
extinction as the afterglow (which can be inferred by demanding that
the temporal and intrinsic spectral slopes, which both depend on the
electron distribution index, $p$, be consistent; see
e.g.\ \citealt{pbr+02}).

To calculate the SN detection limit of our observations, we fit the
model by minimizing $\chi^2$.  The afterglow was not detected in any
of the F450W images, and so we exclude them from our analysis.
Subtracting the host F450W image from our ground-based $g'$ image left
a large residual at the position of the host galaxy (not of the OT).
This poor subtraction is likely due to the filter mis-match, and so we
do not include this point in our analysis.

Our analyses are summarized in Table~\ref{tab:fit}. In short, we find
no evidence for an underlying SN. In order to calculate the formal
limits, we re-fit the data for a range of values of the SN brightness
and computed the probability distribution from the resultant
$\Delta\chi^2$.  As can be seen from Table~\ref{tab:sn}, the least
constraining limit comes from the case where the afterglow evolves in
a wind-stratified medium with the cooling break redward of the optical
band, and even in this case, a SN brighter than $\delta = 1.40$ mag is
excluded at 99.7\% confidence, and a SN as bright as
SN~1998bw ($\delta = 0$ mag) is ruled out at greater than 99.999\%
confidence.

The peak brightness and the time scales for SNe Ib/c are generally
correlated such that fainter SNe may peak earlier \citep{imn+98}.
It may be important to take this into account for our analysis, since
the observations most sensitive to the presence of an underlying SN
are all after the peak.  To do this, we shifted the $UBVRI$ photometry
of the (intrinsically-)fainter Type Ic SN~1994I \citep{rdh+96} to
$z=0.451$, and derived the transformation between the redshifted
SN~1998bw and 1994I light curves using a similar method as
\citet{bkp+02}.  This method is analogous to the ``stretch'' method
for SNe Ia \citep{pag+99}.  If we use this transformation in our model
to transform the redshifted SN~1998bw light curve to the light curve
of a SN fainter than SN~1998bw by $\delta$ magnitudes, then our
least-constraining limit on an underlying SN becomes $\delta = 1.34$
mag fainter than SN~1998bw (at 99.7\% confidence).  The agreement with
the above limit indicates that the uncertainty in our knowledge of the
the light-curve shape and luminosity scaling light-curve is not
important for this analysis.

Leaving aside the SN issue, our fits provide a jet-break time of
approximately 35 days.  From the FREGATE 8 -- 400 keV fluence of $1.5
\times 10^{-5}$ erg cm$^{-2}$, we calculate the $k$-corrected
isotropic-equivalent energy release \citep{bfs01} in the $\gamma$-ray
band, $E_\gamma \sim (1.3 \pm 0.3) \times 10^{52}$ erg.  Applying the
geometric correction from our measurement of the jet break (using the
formulation and normalization of \citealt{fks+01}), we obtain a jet
opening angle of $18^\circ$. Thus the true energy release is $(6.5 \pm
1.6) \times 10^{50}$ erg --- consistent with the clustering of energy
releases around $5 \times 10^{50}$ erg \citep{fks+01}.


\section{Conclusions}
\label{sec:conclusions}

Here we report the search for an underlying SN in the afterglow of
GRB~010921. 
Thanks to the superb photometric stability of HST and the $N\times(N-1)/2$
subtraction technique, we have been able to trace the light curve of
the afterglow of GRB~010921 over two months. The resulting photometry is
unbiased by aperture effects that are so prevalent in simple aperture
and PSF-fitting photometry. We report two results.

First,  we find a jet break time of 35 days, using only optical data.
Second, we find no evidence for an SN.  A SN, if present, must be
fainter than SN~1998bw by $> 1.34$ mag at 99.7\% confidence.  To our
knowledge, to date, this is the most stringent limit for an underlying
SN associated with a cosmologically located GRB.

As noted in \S\ref{sec:introduction}, the collapsar model as currently
understood has little power in predicting the dispersion in the amount
of $^{56}$Ni synthesized as compared to the energy in relativistic
ejecta. Underlying SNe are directly powered by the former whereas the
GRB is powered by the latter. Observations are needed to start mapping
the distribution in these critical explosion parameters.  Progress can
be expected with such observational inputs accompanied by further
refinements in the model. Motivated thus, we summarize in
Table~\ref{tab:previous} the status of SN searches for all
Table~\ref{tab:previous} all known GRBs with
redshift\footnotemark\footnotetext{Beyond a redshift of $\sim$ 1.2,
the distinctive and strong absorption blueward of 4000\AA\ is
redshifted out of the optical bands.  The higher sensitivity of the
optical bands thus favor searches for SNe below this redshift.} less
than 1.2.

The most secure case for an SN is that for GRB~011121
\citep{bkp+02,gsw+02}.  GRB~980326 shows a strong red excess at about
a month but unfortunately a redshift is lacking.  GRB~970228 shows a
less clear excess but benefits from a known redshift. Stated
conservatively, a SN as bright as that of SN~1998bw can be ruled out
in GRB~000911. In all cases, save that of GRBs~980326 and 011121, the
presence of a host with a magnitude comparable to the brightness of
the peak of the SN, makes it difficult to identify an SN component.
As noted in \S\ref{sec:subphot}, ``bumps'' can arise from host
contamination.  Combining HST and ground based measurements (as is the
case for GRB~970228) is prone to considerable errors
(\S\ref{sec:subphot}).

In summary, there is good evidence for an SN comparable in brightness
to SN~1998bw in GRB~011121 \citep{bkp+02}. For GRB~010921, using the
HST observations reported here, we constrain any putative underlying
SN to be 1.34~mag fainter than SN~1998bw.  In the collapsar framework,
this absence could be most readily attributed to the well known
dispersion of the peak luminosity of Type Ib/c SNe.

An alternative possibility is that there may be more than one type of
progenitor for long duration GRBs. Along these lines we note that
\citet{cl00} claim that some afterglows (e.g. GRB~990123) are
incompatible with a $\rho\propto r^{-2}$ inhomogeneous circumburst
distribution whereas other afterglows (e.g. GRBs~970228 and
970508) are better explained by invoking an inhomogeneous circumburst
medium. Progress requires both searches for underlying SNe as well as
characterizing the circumburst medium via modeling of the early-time
afterglow (e.g. GRB~011121, see \citealt{pbr+02}).

Finally, we note that the afterglow of GRB~010921 (and any coincident
SN) was extincted by $A_V^{\rm MW} \approx 0.5$ mag of dust in the
foreground, and $A_V^{\rm host} \approx 1$ mag of dust in the host
galaxy (Table~\ref{tab:fit}). Thus, in the future, using ACS aboard
HST it should be possible to extend SN searches to at least 3 mag
fainter than SN~1998bw, at which point it will be possible to detect
more typical SNe Ib/c coincident with GRBs.

\acknowledgements

We thank Pete Challis for helpful discussions about WFPC2 reduction,
and Megan Novicki and John Tonry for an advance copy of their
$N\times(N-1)/2$ subtraction paper.  SRK and SGD thank NSF for
supporting our ground-based GRB observing program.  BPS and PAP thank
the ARC for supporting Australian GRB research.  Support for Proposal
number HST-GO-08867.01-A was provided by NASA through a grant from
Space Telescope Science Institute, which is operated by the
Association of Universities for Research in Astronomy, Incorporated,
under NASA Contract NAS5-26555.  KH is grateful for support under
grant HST-GO-09180.07-A.


\begin{thebibliography}{}

\bibitem[{Alard}(2000)]{alard00}
{Alard}, C. 2000, \aaps, 144, 363.

\bibitem[{Berger} {\it et al.}\ (2001)]{bdf+01}
{Berger}, E. {\it et al.}\  2001, \apj, 556, 556.

\bibitem[{Berger}, {Kulkarni} \& {Frail}(2001)]{bkf01}
{Berger}, E., {Kulkarni}, S.~R., and {Frail}, D.~A. 2001, apj, 560, 652.

\bibitem[{Bessell}(1990)]{bessell90}
{Bessell}, M.~S. 1990, \pasp, 102, 1181.

\bibitem[{Bj{\" o}rnsson} {\it et al.}\ (2001)]{bhj+01}
{Bj{\" o}rnsson}, G., {Hjorth}, J., {Jakobsson}, P., {Christensen}, L., and
  {Holland}, S. 2001, \apjl, 552, L121.

\bibitem[{Bloom}, {Frail} \& {Sari}(2001)]{bfs01}
{Bloom}, J.~S., {Frail}, D.~A., and {Sari}, R. 2001, \aj, 121, 2879.

\bibitem[{Bloom}, {Kulkarni} \& {Djorgovski}(2002)]{bkd02}
{Bloom}, J.~S., {Kulkarni}, S.~R., and {Djorgovski}, S.~G. 2002, \aj, 123,
  1111.

\bibitem[{Bloom} {\it et al.}\ (1999)]{bkd+99}
{Bloom}, J.~S. {\it et al.}\  1999, Nature, 401, 453.

\bibitem[{Bloom} {\it et al.}\ (2002)]{bkp+02}
{Bloom}, J.~S. {\it et al.}\  2002, \apjl, 572, L45.

\bibitem[{Cardelli}, {Clayton} \& {Mathis}(1989)]{ccm89}
{Cardelli}, J.~A., {Clayton}, G.~C., and {Mathis}, J.~S. 1989, ApJ, 345, 245.

\bibitem[{Castro-Tirado} {\it et al.}\ (2001)]{csg+01}
{Castro-Tirado}, A.~J. {\it et al.}\  2001, \aap, 370, 398.

\bibitem[{Chevalier} \& {Li}(2000)]{cl00}
{Chevalier}, R.~A. and {Li}, Z. 2000, \apj, 536, 195.

\bibitem[{Djorgovski} {\it et al.}\ (2001)a]{dfk+01a}
{Djorgovski}, S.~G., {Frail}, D.~A., {Kulkarni}, S.~R., {Bloom}, J.~S.,
  {Odewahn}, S.~C., and {Diercks}, A. 2001a, \apj, 562, 654.

\bibitem[{Djorgovski} {\it et al.}\ (2001)b]{dkb+01}
{Djorgovski}, S.~G. {\it et al.}\  2001b, in { Gamma-Ray Bursts in the
  Afterglow Era, Proceedings of the International workshop held in Rome, CNR
  headquarters, 17-20 October, 2000. Edited by Enrico Costa, Filippo Frontera,
  and Jens Hjorth. Berlin Heidelberg: Springer}, 218+.

\bibitem[{Dolphin}(2000)]{dolphin00}
{Dolphin}, A.~E. 2000, \pasp, 112, 1397.

\bibitem[{Fitzpatrick} \& {Massa}(1988)]{fm88}
{Fitzpatrick}, E.~L. and {Massa}, D. 1988, \apj, 328, 734.

\bibitem[{Frail} {\it et al.}\ (2002)]{fbm+01}
{Frail}, D.~A. {\it et al.}\  2002, apj, 565, 829.

\bibitem[{Frail} {\it et al.}\ (2001)]{fks+01}
{Frail}, D.~A. {\it et al.}\  2001, ApJ, 562, L55.

\bibitem[{Fruchter} \& {Hook}(2002)]{fh02}
{Fruchter}, A.~S. and {Hook}, R.~N. 2002, \pasp, 114, 144.

\bibitem[{Fukugita} {\it et al.}\ (1996)]{fig+96}
{Fukugita}, M., {Ichikawa}, T., {Gunn}, J.~E., {Doi}, M., {Shimasaku}, K., and
  {Schneider}, D.~P. 1996, \aj, 111, 1748.

\bibitem[{Galama} {\it et al.}\ (2000)]{gtv+00}
{Galama}, T.~J. {\it et al.}\  2000, ApJ, 536, 185.

\bibitem[{Galama} {\it et al.}\ (1998)]{gvv+98}
{Galama}, T.~J. {\it et al.}\  1998, Nature, 395, 670.

\bibitem[{Galama} \& {Wijers}(2001)]{gw01}
{Galama}, T.~J. and {Wijers}, R.~A.~M.~J. 2001, \apjl, 549, L209.

\bibitem[{Garnavich} {\it et al.}\ (2002)]{gsw+02}
{Garnavich}, P. {\it et al.}\  2002, ApJ (submitted), astro-ph/0204234.

\bibitem[{Garnavich} {\it et al.}\ (2000)]{gjp+00}
{Garnavich}, P.~M., {Jha}, S., {Pahre}, M.~A., {Stanek}, K.~Z., {Kirshner},
  R.~P., {Garcia}, M.~R., {Szentgyorgyi}, A.~H., and {Tonry}, J.~L. 2000, \apj,
  543, 61.

\bibitem[{Halpern} {\it et al.}\ (2000)]{hum+00}
{Halpern}, J.~P. {\it et al.}\  2000, \apj, 543, 697.

\bibitem[{Harrison} {\it et al.}\ (2001)]{hys+01}
{Harrison}, F.~A. {\it et al.}\  2001, \apj, 559, 123.

\bibitem[{Henden}(2001)]{henden01}
{Henden}, A. 2001, {GCN} Circular 1100.

\bibitem[{Hjorth} {\it et al.}\ (1999)]{hpj+99}
{Hjorth}, J., {Pedersen}, H., {Jaunsen}, A.~O., and {Andersen}, M.~I. 1999,
  \aaps, 138, 461.

\bibitem[{Holland} {\it et al.}\ (2001)]{hfh+01}
{Holland}, S. {\it et al.}\  2001, \aap, 371, 52.

\bibitem[Hurley {\it et al.}\ (2001)]{hurley+01}
Hurley, K., Cline, T., Frontera, F., Montanari, E., Guidorzi, C., Feroci, M.,
  and the HETE Science~Team 2001, {GCN} Circular 1097.

\bibitem[{Iwamoto} {\it et al.}\ (1998)]{imn+98}
{Iwamoto}, K. {\it et al.}\  1998, \nat, 395, 672.

\bibitem[Kulkarni {\it et al.}\ (1998)]{kfw+98}
Kulkarni, S.~R. {\it et al.}\  1998, Nature, 395, 663.

\bibitem[{Lazzati} {\it et al.}\ (2001)]{lcg+01}
{Lazzati}, D. {\it et al.}\  2001, \aap, 378, 996.

\bibitem[{MacFadyen} \& {Woosley}(1999)]{mw+99}
{MacFadyen}, A.~I. and {Woosley}, S.~E. 1999, \apj, 524, 262.

\bibitem[{Park} {\it et al.}\ (2002)]{pwh+02}
{Park}, H.~S. {\it et al.}\  2002, \apjl, 571, L131.

\bibitem[{Perlmutter} {\it et al.}\ (1999)]{pag+99}
{Perlmutter}, S. {\it et al.}\  1999, \apj, 517, 565.

\bibitem[{Piro} {\it et al.}\ (2002)]{pfg+02}
{Piro}, L. {\it et al.}\  2002, ApJ (in press), astro-ph/0201282.

\bibitem[{Piro} {\it et al.}\ (2000)]{pgg+00}
{Piro}, L. {\it et al.}\  2000, Science, 290, 955.

\bibitem[{Price} {\it et al.}\ (2002a)]{pbr+02}
{Price}, P.~A. {\it et al.}\  2002a, \apj, 572, L51.

\bibitem[{Price} {\it et al.}\ (2001)]{phg+01}
{Price}, P.~A. {\it et al.}\  2001, \apjl, 549, L7.

\bibitem[{Price} {\it et al.}\ (2002b)]{pkb+02}
{Price}, P.~A. {\it et al.}\  2002b, \apj, 571, L121.

\bibitem[{Reeves} {\it et al.}\ (2002)]{rwo+02}
{Reeves}, J.~N. {\it et al.}\  2002, \nat, 416, 512.

\bibitem[{Reichart}(1999)]{reichart99}
{Reichart}, D.~E. 1999, \apj, 521, L111.

\bibitem[{Reichart}(2001)]{reichart01}
{Reichart}, D.~E. 2001, ApJ, 553, 235.

\bibitem[{Richmond} {\it et al.}\ (1996)]{rdh+96}
{Richmond}, M.~W. {\it et al.}\  1996, \aj, 111, 327.

\bibitem[{Ricker} {\it et al.}\ (2002)]{rhl+02}
{Ricker}, G. {\it et al.}\  2002, \apj, 571, L127.

\bibitem[{Sari}, {Piran} \& {Halpern}(1999)]{sph99}
{Sari}, R., {Piran}, T., and {Halpern}, J.~P. 1999, ApJ, 519, L17.

\bibitem[{Sari}, {Piran} \& {Narayan}(1998)]{spn98}
{Sari}, R., {Piran}, T., and {Narayan}, R. 1998, ApJ, 497, L17.

\bibitem[{Schmidt} {\it et al.}\ (1998)]{ssp+98}
{Schmidt}, B.~P. {\it et al.}\  1998, \apj, 507, 46.

\bibitem[{Sokolov}(2001)]{sokolov01}
{Sokolov}, V.~V. 2001, in { Gamma-Ray Bursts in the Afterglow Era, Proceedings
  of the International workshop held in Rome, CNR headquarters, 17-20 October,
  2000. Edited by Enrico Costa, Filippo Frontera, and Jens Hjorth. Berlin
  Heidelberg: Springer}, 136.

\bibitem[{Vietri} \& {Stella}(1999)]{vs99}
{Vietri}, M. and {Stella}, L. 1999, \apjl, 527, L43.

\bibitem[Woosley(1993)]{woo93}
Woosley, S.~E. 1993, ApJ, 405, 273.

\end{thebibliography}

\clearpage

\begin{deluxetable}{ccccc}
\footnotesize
\tablecolumns{5}
\tablewidth{0pt}
\tablecaption{\label{tab:hst}HST+WFPC2 observations of GRB~010921}
\tablehead{\colhead{Date (2001, UT)} & \colhead{Filter} & \colhead{Flux ($\mu$Jy)} }
\startdata
Oct 26.731	&      F450W	&   -0.031 $\pm$ 0.022	\\
Nov 06.956	&      F450W	&   0.001 $\pm$ 0.028 	\\
Nov 24.990	&      F450W	&   0.067 $\pm$ 0.029 	\\
Oct 26.791	&      F555W	&   0.157 $\pm$ 0.015 	\\
Nov 07.015	&      F555W	&   0.087 $\pm$ 0.017 	\\
Nov 25.121	&      F555W	&   0.063 $\pm$ 0.018 	\\
Oct 26.859	&      F702W	&   0.231 $\pm$ 0.013 	\\
Nov 07.149	&      F702W	&   0.096 $\pm$ 0.015 	\\
Nov 25.203	&      F702W	&   0.045 $\pm$ 0.015 	\\
Oct 26.932	&      F814W	&   0.433 $\pm$ 0.024 	\\
Nov 08.359	&      F814W	&   0.209 $\pm$ 0.024 	\\
Nov 25.621	&      F814W	&   -0.003 $\pm$ 0.025	\\
Oct 26.992	&      F850LP	&  0.471 $\pm$ 0.092 	\\
Nov 08.418	&      F850LP	&  0.207 $\pm$ 0.088 	\\
Nov 25.687	&      F850LP	&  0.030 $\pm$ 0.096 	\\
\enddata
\tablecomments{These host-subtracted measurements have not been
corrected for Galactic extinction, and are all made under the
assumption that the flux of the OT on 2001 Dec 21 is zero or
negligible, $<< 0.01 \mu$Jy.}
\end{deluxetable}

\clearpage

\begin{deluxetable}{ccccc}
\footnotesize
\tablecolumns{5}
\tablewidth{0pt}
\tablecaption{\label{tab:ground}Re-analysis of ground-based observations
of GRB~010921}
\tablehead{\colhead{Date (2001, UT)} & \colhead{Filter} & \colhead{Flux ($\mu$Jy)} & \colhead{Telescope} }
\startdata
Oct 19.178	&      $g'$	&	0.671 $\pm$ 0.097	&	P200	\\
Sep 22.144	&      $r'$	&	46.104 $\pm$ 0.722	&	P200	\\
Sep 22.148	&      $r'$	&	44.995 $\pm$ 0.661	&	P200	\\
Sep 27.354	&      $r'$	&	2.13 $\pm$ 1.223	&  	P200	\\
Oct 17.145	&      $r'$	&	0.086 $\pm$ 0.379	& 	P200	\\
Oct 18.088	&      $r'$	&	0.189 $\pm$ 0.382	& 	P200	\\
Oct 19.109	&      $r'$	&	0.256 $\pm$ 0.285	& 	P200	\\
Oct 17.165	&      $i'$	&	0.560 $\pm$ 0.197	& 	P200	\\
Oct 18.110	&      $i'$	&	0.523 $\pm$ 0.191	& 	P200	\\
Oct 19.130	&      $i'$	&	0.649 $\pm$ 0.153	& 	P200	\\
Oct 19.149	&      $z'$	&	1.293 $\pm$ 4.273	& 	P200	\\
Sep 22.3038	&     $B$	&	11.319 $\pm$ 0.981	&	NOFS1.0	\\
Oct 19.253	&      $B$	&	0.623 $\pm$ 0.675	& 	P60	\\
Sep 22.2976	&     $V$	&	24.727 $\pm$ 1.078	&	NOFS1.0	\\
Oct 19.206	&      $V$	&	0.229 $\pm$ 0.720	& 	P60	\\
Sep 22.2930	&     $R$	&	39.116 $\pm$ 5.072	&	NOFS1.0	\\
Sep 22.3210	&     $R$	&	36.135 $\pm$ 4.486	&	NOFS1.0	\\
Oct 19.272	&      $R$	&	0.916 $\pm$ 4.284	& 	P60	\\
Nov 17.151	&      $R$	&	0.470 $\pm$ 4.238	& 	NOFS1.0	\\
Sep 22.2893	&     $I$	&	84.688 $\pm$ 5.778	&	NOFS1.0	\\
Sep 22.795	&	$I$	&	40.277 $\pm$ 3.950	&	TS	\\
Sep 22.825	&	$I$	&	47.281 $\pm$ 7.230	&	TS	\\
Sep 22.878	&	$I$	&	50.926 $\pm$ 3.671	&	TS	\\
Sep 22.954	&	$I$	&	41.321 $\pm$ 3.636	&	TS	\\
Nov 17.093	&      $I$	&	1.229 $\pm$ 1.057	& 	NOFS1.0	\\
\enddata
\tablecomments{These host-subtracted measurements have not been
corrected for Galactic or host extinction, and are all made under the
assumption that the flux of the OT on 2001 Dec 21 is zero or
negligible.  Zero-points were set from the star at coordinates
R.A. = 22$^{\rm h}$56$^{\rm m}$00$^{\rm s}$.21 Dec =
40$^{\circ}$54$'$58$''$.0 with $B = 21.248$ mag, $V = 20.230$ mag,
$R = 19.699$ mag and $I = 19.132$ mag, accurate to better than
3 percent \citep{henden01}.  Telescopes are: P200 --- Hale Palomar 200-inch;
NOFS1.0 --- USNO Flagstaff Station 1.0-metre; P60 --- Palomar 60-inch;
TS --- Tautenburg Schmidt.  NOFS1.0 observations of Sep 22, and all P200
and P60 observations were presented in \citet{pkb+02}; NOFS1.0 and
TS observations were presented in \citet{pwh+02}.}
\end{deluxetable}

\clearpage

\begin{deluxetable}{lrrrr}
\footnotesize
\tablecolumns{6}
\tablewidth{0pt}
\tablecaption{\label{tab:fit} Best-fit afterglow models  }
\tablehead{\colhead{Model} & \colhead{$p$} & \colhead{$t_{\rm jet}$/days} & \colhead{$A_V^{\rm host}$/mag} & \colhead{$\chi^2$} }
\startdata
ISM/Wind,B	&	2.67 $\pm$ 0.06	&	33.0 $\pm$ 6.5	&	0.95 $\pm$ 0.08	&	19.9	\\	
ISM,R		&	3.03 $\pm$ 0.04	&	37.5 $\pm$ 4.9	&	1.16 $\pm$ 0.07	&	19.2	\\
Wind,R		&	2.33 $\pm$ 0.10	&	30.3 $\pm$ 9.5	&	1.35 $\pm$ 0.08	&	23.1	\\
\enddata
\tablecomments{The best-fit afterglow parameters from fitting a standard
afterglow model with host extinction and no SN.  Each fit had 32 degrees
of freedom.  ISM models refer to afterglow evolution in an homogeneous ISM.
Wind models refer to afterglow evolution in a wind-stratified ($r^{-2}$)
medium.  R and B refer to the location of the cooling break relative to
the optical bands (redward and blueward, respectively).  If the cooling
break is blueward of the optical bands, then the ISM and wind models
both have the same form.}
\end{deluxetable}

\clearpage

\begin{deluxetable}{lrrr}
\footnotesize
\tablecolumns{6}
\tablewidth{0pt}
\tablecaption{\label{tab:sn} Maximum Allowed Brightness of a SN underlying GRB~010921}
\tablehead{\colhead{Significance} & \colhead{ISM/Wind,B} & \colhead{ISM,R} & \colhead{Wind,R} }
\startdata
1$\sigma$ (68.3\%)	&	3.17 (2.89)	&	2.76 (2.61)	&	1.80 (2.39)	\\
2$\sigma$ (95.4\%)	&	2.29 (2.17)	&	1.98 (1.94)	&	1.55 (1.70)	\\
3$\sigma$ (99.7\%)	&	1.80 (1.73)	&	1.87 (1.57)	&	1.40 (1.34)	\\
\enddata
\tablecomments{For each model and significance level, we list the magnitude
relative to SN~1998bw of the faintest SN detectable by the observations.
In brackets, we include the magnitude relative to SN~1998bw of a `generic
SN'.  Model descriptions are the same as in Table~\ref{tab:fit}.}
\end{deluxetable}

\clearpage

\begin{deluxetable}{lrcrrll}
\rotate
\footnotesize
\tablecolumns{5}
\tablewidth{0pt}
\tablecaption{\label{tab:previous}Other GRB-SNe}
\tablehead{\colhead{GRB} & \colhead{$z$} & \colhead{Band} & \colhead{SN (mag)} & \colhead{Host (mag)} & \colhead{Ground/HST?} & \colhead{Comments, ref} }
\startdata
970228	& 0.695 & $R$ &	25.5 &	25.2 &	Both &	Plausible but aperture, color effects with HST. (1) \\
970508	& 0.835	& $I$ &	23.6 &	24.0 &	Ground&	Aperture effects. (2)\\
980326	& ???	& $R$ &	25 &	$<$ 27 & Ground& Plausible. (3) \\
980613	& 1.096	& $R$ &	\ldots&	24.0 &	Ground&	Faint afterglow, no search. (4)\\
980703	& 0.966	& $R$ &	24 &	22.6 &	Ground&	Consistent with no SN. (5)\\
990705	& 0.840	& $R$ & \ldots&	22.8 &	Ground&	No search. \\
990712	& 0.433	& $V$ &	23.8 &	21.2 &	Ground&	Aperture effects? (6) \\
991208	& 0.706	& $R$ &	23.9 &	24.4 &	Ground&	Bad afterglow fit. (7) \\
991216	& 1.020	& $R$ &	\ldots&	24.85 &	Ground&	No search, consistent with no SN. (8) \\
000418	& 1.119	& $R$ &	\ldots&	23.8 &	Ground&	Consistent with no SN. (9) \\
000911	& 1.058	& $I$ &	24.7 &	24.4 &	Ground&	2$\sigma$ detection, SN $\sim$ 0.9 $\pm$ 0.3 $\times$ SN1998bw. (10) \\
011121	& 0.365	& $R$ &	23 &	26 &	HST &	Secure. (11) \\
\enddata
\tablecomments{All GRBs with optical afterglows at $z < 1.2$,
excepting GRB~980425 (SN~1998bw) and GRB~010921 (this study).  
References: 1:
\citet{reichart99,gtv+00}; 2: \citet{sokolov01}; 3: \citet{bkd+99}; 4:
\citet{hpj+99}; 5: \citet{hfh+01}; 6: \citet{bhj+01}; 7:
\citet{csg+01}; 8: \citet{gjp+00,hum+00}; 9: \citet{bdf+01}; 10:
\citet{lcg+01}; 11: \citet{bkp+02}.  Host magnitudes and redshifts
were also compiled from \citet{dkb+01}.}
\end{deluxetable}

\clearpage

\end{document}